\documentclass[twocolumn,%
 reprint,
 amsmath,amssymb,
 aps,
prd,
]{revtex4-1}

\usepackage[usestackEOL]{stackengine}
\usepackage{graphicx}
\usepackage{dcolumn}
\usepackage{bm}

\usepackage{enumerate}
\usepackage{nicefrac}
\usepackage{lipsum}

\newcommand{\bea}{\begin{eqnarray}}
\newcommand{\eea}{\end{eqnarray}}

\usepackage{centernot}
\usepackage{color}
\newcommand{\be}{\begin{equation}}
\newcommand{\ee}{\end{equation}}

\makeatletter
\let\Hy@backout\@gobble
\makeatother

\begin{document}


\title{Quantum Chameleons
}

\author{Philippe Brax,$\,^a$ Sylvain Fichet$^{\,b,c}$}
\email{philippe.brax@ipht.fr, sylvain@ift.unesp.br}

\affiliation{%
$^a$Institut de Physique Th\'{e}orique, Universit\'e Paris-Saclay, CEA, CNRS, F-91191 Gif/Yvette Cedex, France \\
$^b$Walter Burke Institute for Theoretical Physics, California Institute of Technology, Pasadena, CA
91125, California, USA \\
$^c$ICTP-SAIFR \& IFT-UNESP, R. Dr. Bento Teobaldo Ferraz 271, S\~ao Paulo, Brazil
}

\begin{abstract}

We initiate a quantum treatment of chameleon-like particles, deriving classical and quantum forces directly from the path integral. It is found that the quantum force  can potentially dominate the classical one by many orders of magnitude. We calculate the quantum chameleon pressure between infinite plates, which is found to interpolate between the Casimir and the integrated Casimir-Polder pressures, respectively in the limits of full screening and no screening. To this end we calculate the chameleon propagator  in the presence of  an arbitrary number of one-dimensional layers of material. For the E\"ot-Wash experiment, the five-layer propagator is used to take into account the  intermediate shielding sheet, and it is found that the presence of the sheet enhances the quantum pressure by two orders of magnitude. As an example of implication, we show
that in both the standard chameleon and symmetron models,  large and previously unconstrained regions of the parameter space are excluded once the quantum pressure is taken into account.


\end{abstract}

\maketitle

\section{Introduction}


A wealth of Dark Energy models involve a scalar field  with an extremely low mass which plays a role on cosmological scales~\cite{Brax:2013ida}, explaining the accelerated expansion of the Universe. At shorter distances, such as the Solar System  scale, screening  mechanisms must take place to suppress the long-range force induced by the new scalar, since such scenario would be otherwise  excluded by  stringent experimental tests~\cite{Bertotti:2003rm,Williams:2012nc}.
Screening mechanisms of the long-range force can naturally occur as a result of the scalar coupling to matter.  Indeed, whenever the local matter density  is high enough with respect to the other scales of the problem, the pro\-perties of the scalar (mass or couplings) tend to change in the local environment and, typically, the scalar  tends to get invisible where one could observe it~\cite{Khoury:2003aq,Khoury:2003rn,Brax:2004qh,Brax:2010kv}. We will refer to any scalar with such property as a \textit{chameleon-like} field (and will sometimes use only ``chameleon'' for short).
For instance for the original chameleon model  the mass of the field increases in dense environments whilst in the symmetron model \cite{Pietroni:2005pv,Olive:2007aj,Hinterbichler:2010es}  screening occurs  as the coupling to matter decreases with an increasing matter density.
The existence of chameleon-like fields can be tested by laboratory experiments, for instance by neutrons \cite{Brax:2011hb,Jenke:2014yel,Lemmel:2015kwa, Cronenberg:2018qxf} or atomic spectroscopy \cite{Burrage:2015lya,Jaffe:2016fsh}. The pressure between two parallel plates is also suited to test the potential presence of chameleons \cite{Brax:2007vm,Brax:2010xx}
which could become within reach in the near future \cite{Almasi:2015zpa}.
%

The effects of chameleon-like fields are typically treated in a classical approximation. However at short enough distances -- such as the submicron scale in the  E\"ot-Wash experiment \cite{Adelberger:2006dh}, a quantum treatment of the chameleon mechanism becomes mandatory. In this work we develop the formalism to describe ``quantum chameleons''  and present some of its consequences. { The formalism also sheds new light on the quantum field theory calculation of the Casimir pressure in its various regimes, providing for instance the general quantum pressure interpolating between the  Casimir and the Casimir-Polder limits.


The paper is arranged as follows. In section \ref{sec:force} we calculate the force between bodies due to a chameleon-like field at the classical and quantum levels. In section \ref{sec:plate} we restrict ourselves to the quantum chameleon force between parallel plates. We then apply our results to obtain constraints from fifth force experiments
 in section \ref{sec:eot}.
 We conclude in Sec.~\ref{se:conc} and  more details on    the  Feynman propagators and the Casimir-Polder force are given in the Appendix.  }


%
%

\section{Chameleon forces from the path integral}
\label{sec:force}

{Our   focus in this work is on scalar-tensor theories with a conformal coupling between the scalar field and matter \cite{Damour:1992we}}. In the Einstein frame this translates into  the  general chameleon  Lagrangian {
\be
{\cal L}[\phi,J]= \frac{1}{2}(\partial_\mu \phi)^2 - V(\phi) -A(\phi)J \,,
\ee
where gravity is described by its Einstein-Hilbert Lagrangian, $V(\phi)$ is the interaction potential of the scalar field, and $J$ is a  source} {representing the density of pressure-less matter in the setting}.    The term  $A(\phi)J$  describes how the chameleon couples to the source~\footnote{Our conventions follow the ones of \cite{Peskin:257493}}.
In this works the source  $J$ is  considered as a matter density, {i.e.} $J(x)=\rho(x)$, {which} is assumed to be static and vanishes at infinity.  In general $J$, $\phi$, $A(\phi)$ depend on space. Both $V$ and $A$ can contain higher dimensional operators, in which case the Lagrangian describes a low-energy effective field theory (EFT) {valid for distances greater than some UV cutoff scale}.  In the following we define the effective potential
\be
V_J=V+AJ \, .
\ee
The source is assumed to depend  on an external parameter $L$, to be understood typically  as measuring the distance between two objects, $L=|L^i|$. Our goal is to study how the quantum system reacts when changing $L$.

{  We are interested in calculating the energy of a configuration involving several objects acting as sources, all described by the distribution $J$. The energy involves both the classical energy due to the
classical field configuration between the bodies and the quantum fluctuations around the classical value.
The energy of the set of objects can be obtained by integrating over the scalar field. This can be performed using a path integral $Z[J]$ whose source term is the $J$ distribution representing the objects.  The relevant information is contained in the generating {functional} of connected correlators, given by
\be
W[J]=i\log Z[J] \,, \quad Z[J]=\int {\cal D}\phi e^{i \int d^4 x  {\cal L}[\phi,J]} \,.
\ee
This is the Lorentzian analog to the free-energy in Euclidian space (see \textit{e.g} \cite{Peskin:257493}). When the source is static, $W[J]$  involves only a potential energy which is given by
\be
 E[J]=\frac{W[J]}{T}\,
\ee
where $T=\int dt$ is the integral over time.  We will work with  $T=1$ conventionally and refer  to the potential $E[J]$  as the vacuum energy.
Our conventions regarding spacetime coordinates are $x^\mu=(t,x^i)$, $d^4x\equiv dt d^3 x$, $d^3x\equiv dx^1 dx^2 dx^3$.

}

\begin{figure}
\includegraphics[width=8.5 cm,trim={2.2cm 6cm 3.2cm 2cm},clip]{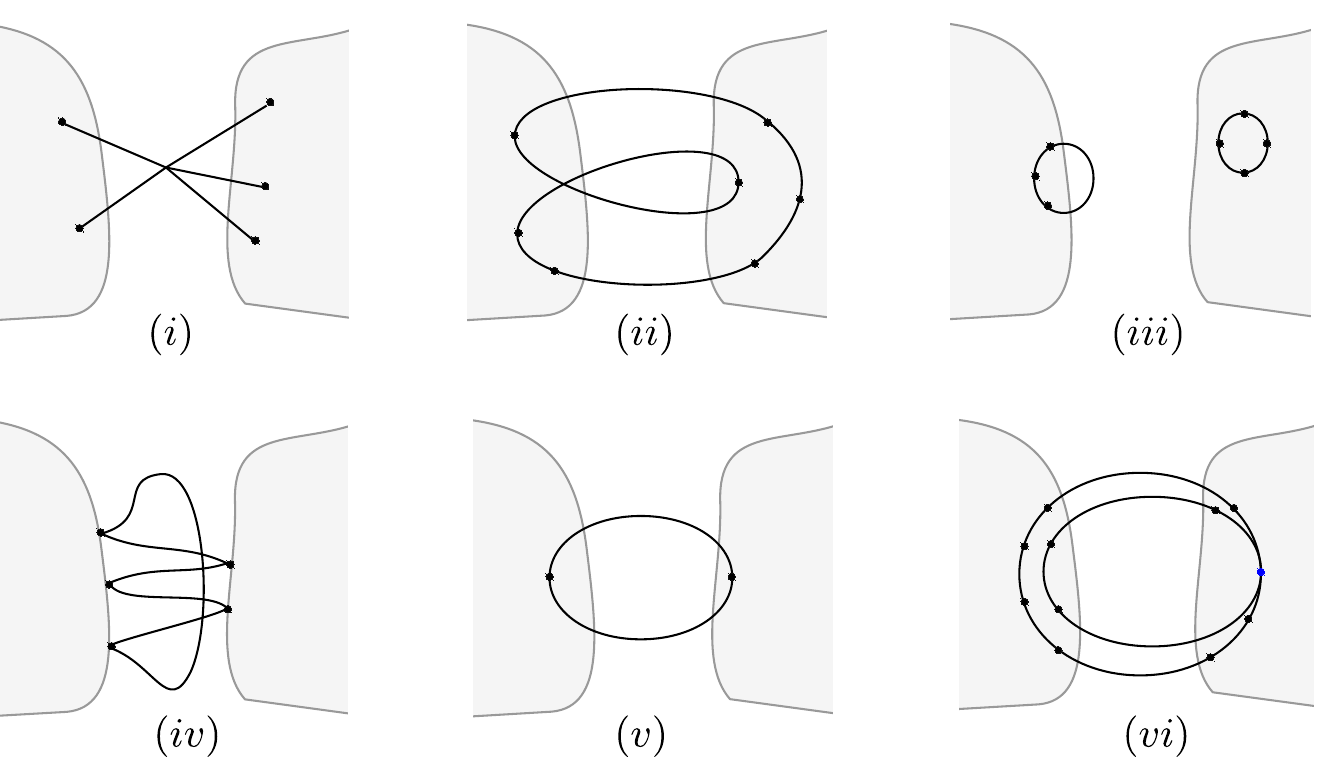}\\
\caption{Sample Feynman diagrams for a chameleon field in the presence of two extended sources. \textit{(i)}:  A generic classical contribution. \textit{(ii)}: A generic 1-loop contribution.
\textit{(iii)}: Tadpoles. \textit{(iv)}:  Casimir (Strong coupling to sources.)
\textit{(v)}:Casimir-Polder (Weak coupling to sources.)
\textit{(vi)}: A two-loop contribution from higher-dimensional coupling to matter.
\label{fig:Diags}
}
\end{figure}

{

All the information about the force (or pressure) that one source induces on another one in the presence of the chameleon-like field is contained in the variation of the va\-cuum energy with respect to $L$.  This variation is given by
\be
\partial_L E[J]=  \frac{\int {\cal D}\phi \int d^3x \partial_L J  A(\phi) e^{i \int d^4 x  {\cal L}[\phi,J]} }{\int {\cal D}\phi e^{i \int d^4 x  {\cal L}[\phi,J]}}\, .
\ee
{As the numerator is simply a normalising constant, this can be rewritten as
\be
\partial_L E[J]= \int d^3x\, \partial_L J \left(\frac{\int {\cal D}\phi A(\phi) e^{i \int d^4 x  {\cal L}[\phi,J]} }{ \int{\cal D}\phi e^{i \int d^4 x  {\cal L}[\phi,J]}}\right )
\ee
where we have commuted the integrals and extracted the factor $\int d^3x \,  \partial_L J$ from the functional integral as this is a field independent term.
The term in brackets can be seen as  the averaged value of $A$ over all the field configurations of $\phi$, leading to the general expression for the force
\be
\partial_L E[J] =  \int d^3 x \,\partial_L J\, \langle A \rangle_J  \,,
\label{eq:delE_simple}
\ee
where
\be
\langle A \rangle_J =\frac{\int {\cal D}\phi A(\phi) e^{i \int d^4 x  {\cal L}[\phi,J]} }{ \int{\cal D}\phi e^{i \int d^4 x  {\cal L}[\phi,J]}}
\label{eq:Aav}
\ee
{{  depends  on the sources $J$. }

{{  Notice that in the functional integral defining $E[J]$, only the source term depends on $L$ and therefore the variation with respect to $L$ is only operative on the source $J$. This is the reason why the variation of the vacuum energy Eq.~\eqref{eq:delE_simple} only involves $\partial_L J$. {From a functional derivative viewpoint, $\langle A \rangle_J$ is  given by  $\langle A \rangle_J = \delta E[J] / \delta J$, \textit{i.e.}
Eq.~\eqref{eq:delE_simple} is simply an elaborate version of the chain rule.
}
Namely, the  formula  decouples the quantum average leading to $\langle A \rangle_J \ $ and the change in the positions of the bodies captured by $\partial_L J$.
In a sense, Eq.~\eqref{eq:delE_simple} is deceptively simple as the calculation of the quantum  average $\langle A \rangle_J$ is highly non-trivial and can in general be evaluated only in  the loop expansion of quantum field theory. }}

To go further, let us thus perform the $\hbar$  expansion. Writing the chameleon field as $\phi=\phi_{\rm cl}+\eta$ where $\eta$ represents the quantum fluctuations around the classical field $\phi_{\rm cl}$, we find the first two terms in the $\hbar$ expansion to be
\begin{align}
\partial_L E &=  \int d^3 x  \partial_L J  A(\phi_{\rm cl})
 + \frac{1}{2} \int d^3 x \partial_L J A'' \Delta_J(x,x) +O(\hbar^2) \nonumber\\
 &= F_{\rm cl} + F_{\rm quant}  +O(\hbar^2) \,,
   \label{eq:delE_full}
\end{align}
\normalsize
where $\Delta_J(x,x')$ is the Feynman propagator of the fluctuation, which satisfies the equation of motion
\be
(\partial^2_x+V''+A''J)\Delta_J(x,x')=-i\delta^{(4)}(x-x')\,,
\label{eq:Delta_prop}
\ee
where $V''=\frac{d^2V}{d\phi^2}\vert_{\phi= \phi_{\rm cl}}$, $A''=\frac{d^2 A}{d\phi^2}\vert_{\phi= \phi_{\rm cl}}$.
The convention adopted here is that   $F>0$ for an  attractive force.
{The Feynman prescription will be specified below and in appendix \ref{app:A}.}
An efficient way to obtain the general result Eq.~\eqref{eq:delE_full} is to expand $\phi$ in Eq.~\eqref{eq:Aav}, which gives  $\langle A(\phi)\rangle=A(\phi_{\rm cl})+\frac{1}{2}A''(\phi_{\rm cl})\langle \eta(x)^2\rangle +O(\hbar^2)$, and realise that $\langle\eta(x)^2\rangle$ is the connected correlator of $\eta$ in presence of the source term, \textit{i.e.} $\Delta_J(x,x)$. A more pedestrian derivation using the 1-loop functional determinant will be described further below and in App.~\ref{app:0}.


{

The first term in Eq.~\eqref{eq:delE_full} is  the \textit{classical} force, pictured in Fig.~\ref{fig:Diags}\textit{i}.
This term does not involve relativistic retardation and only requires to solve the background equation of motion.
When for instance $A(\phi)=y \phi$, $V(\phi)=\frac{1}{2}m^2\phi^2$, one recovers exactly the Yukawa force
  (note  one  can use  $\langle \phi \rangle(x)= y \int d^4x' i\Delta(x,x')  J(x') $) giving  the same result as in \cite{Brax:2014zta}.

One can also notice that when writing explicitly the source as describing two bodies
 $a$ and $b$ in the form  $J(x^i)=J_a(x^i-L^i)+J_b(x^i)$,  using $\partial_{L^i} J_a= - \partial_i J_a$, and using the fact that the bodies vanish at infinity, we obtain
 \be
F_i= \int  d^3 x \,   \rho_a( x^i)   \partial_{i} A(\phi_{\rm cl}) \, \label{eq:Fclass}
\ee
after integrating by part. This matches the classical result used in \cite{Brax:2014zta} and used to calculate classical forces in scalar-tensor theories.

%
%
%


The second term in Eq.~\eqref{eq:delE_full} is  the \textit{quantum} force {at one-loop order}, pictured in Fig.~\ref{fig:Diags}\textit{ii}.
One method to obtain it is to use the explicit evaluation of $E[J]$, which contains the 1-loop functional determinant (see \textit{e.g} \cite{Peskin:257493})
\begin{align}
 E[J] & \supset    -\frac{i}{2} {\rm Tr}\log (\partial^2+V''+ A''J) \label{eq:Trlog} \\
&=\frac{i}{2}\left( \sum_{n=1}^\infty \frac{1}{n} {\rm Tr} \left( - \frac{A'' J}{\partial^2 + V''} \right)^n -{\rm Tr}\log (\partial^2+V'')  \right) . \nonumber
\end{align}
The variation of $E[J]$ with respect to $L$ is detailed in App.~\ref{app:0} and gives
\be
\partial_L E[J]\supset \frac{i} {2} \int d^3 x A'' \partial_L J (-i)\Delta_J (x,x) +O(\hbar^2)
\label{eq:EJvar}
\ee
where the quantity
\be
\Delta_J (x,x')= -i(\partial^2 +V'')^{-1} \sum_{n=0}^\infty  {\rm Tr} \left( -\frac{A'' J}{\partial^2 + V''} \right)^n \label{eq:Deltageom}
\ee
has appeared, which is  precisely the geometric series representation of $\Delta_J$ satisfying Eq.~(\ref{eq:Delta_prop}).
The result Eq.~\eqref{eq:EJvar} reproduces the quantum force formula given in Eq.~\eqref{eq:delE_full}.

The vacuum energy (in)famously contains infinities which usually have to be subtracted by hand (see \textit{e.g.} \cite{Weinberg:1988cp,Milton:2002vm}). In our approach all divergences automatically disappear thanks to the $\partial_L$ as they are $L$-independent, as should be the case as  $\partial_L E[J]$ is an observable.
Indeed, in the   functional determinant,  the $\partial_L$ removes all diagrams which do not link a source to the other one, \textit{i.e.} the ``tadpole'' diagrams of the extended sources, pictured in Fig.~\ref{fig:Diags}\textit{iii}.
 Thus in Eq.~\eqref{eq:delE_full} the infinite part of $\Delta(x,x)$ (which is $L$-independent) does not contribute and one can readily use its finite part $\Delta^{\rm fin}(x,x)$.} More details are given in App.~\ref{app:A}.

\subsection{An aside: Boundary integral representation}

Before discussing further the properties of the chameleon quantum force, it is worth pointing out another representation which applies to the general force Eq.~\eqref{eq:delE_full}, or to any of the term of the $\hbar$ expansion separately. The following formalism applies to bodies with sharp boundaries, whose volumes $V_{a,b}$ can be described in the form
\be
f_{a}(x^i - L^i) \geq 0\,,\quad f_{b}(x^i ) \geq 0 \,.
\ee
The source $a$ can generically be modeled as
\be
J_{a}( x^i)= \rho_{a}( x^i) \Theta( f_{a}(x^i-  L^i))
\ee
with a space-varying density $\rho_{a}( x^i)$, and similarly for $b$.
Such modeling of the sources applies  to essentially all physically relevant situations.

Let us remark  that the variation of  source $a$ with respect to $L^i$ gives
\be
\partial_{L^i} J_a= - \rho_a ( x^i)\partial_{i} f_a(x^i- L^i) \delta( f_a(x^i- L^i))\,. \label{eq:varJa}
\ee
Using Eq.~\eqref{eq:varJa}, the force is given by an integral over  the boundary $S$ of $V_a$. Parametrizing the boundary manifold using coordinates $\xi_{\alpha}$ $(\alpha=1,2)$ with induced metric $g_{\alpha\beta}$, we have
\be
F^i= - \int_S  d^2\xi \sqrt{g} \, n^i \, \rho( x^i) \langle A \rangle_J
\label{eq:claszbound}
\ee
where
\be
n^i = \frac{\partial_{i} f_a(x^i- L^i)}{|\partial_{i} f_a(x^i- L^i)|}\,
\ee
is   the unit vector normal to $S$ and oriented inwards $V_a$.

To study the modulus of the force, let us chose a coordinate system such that $L^i=(0,0,L^z)$ without loss of generality. In these coordinates the force modulus is given by
\be
|F|=F^z= - \int_S  d^2\xi \sqrt{g} \, n^z \, \rho( x^i) \langle A \rangle_J
\label{eq:clasz}
\ee
 Since the volume is closed,  $n^z$ can take both signs. This  naturally splits the integral into a positive contribution from $S_+=(\xi^\alpha|n^z\geq 0)$ and a negative contribution from  $S_-=(\xi^\alpha|n^z<0)$. The presence of these two opposite-sign contributions in the boundary integral is helpful to understand how infinite contributions cancel in the quantum case. This will be explicitly illustrated in the case of plates in Sec.~\ref{sec:plate}.

\subsection{The chameleon quantum force}

Computing the quantum force (Eq.~\eqref{eq:delE_full}) requires to calculate the $\Delta_J$ propagator. However  some important general properties can be deduced  prior to any calculation. Whenever the source term $A''J$ is large with respect to other scales involved in the interaction potential, the Green's function  should vanish (\textit{i.e.} be ``screened'') inside the source and vanish at its surface, as illustrated in Fig.~\ref{fig:Diags}\textit{iv}, see appendix \ref{app:A}.  These are precisely the conditions for the standard Casimir effect.

In the opposite limit, when the coupling to the source $A'' J$  can be treated perturbatively, the functional determinant Eq.~\eqref{eq:Trlog} can be truncated at quadratic order, in which case it is the limit of no screening where the force is
\begin{align}
F_{ \rm quant}=  & \frac{i}{2} \int d^3x\int d^4x'  \\ \nonumber
& A''(x) \partial_L J(x)  \Delta_0(x,x') A''(x')J(x') \Delta_0(x',x) \,.
\end{align}
This corresponds to the bubble diagram shown in Fig.~\ref{fig:Diags}\textit{v}, which is precisely  the Casimir-Polder force integrated over extended sources, see appendix \ref{app:B}.
{ For point sources $ J(x)=\delta^{(3)}(x^i)+\delta^{(3)}(x^i-L^i)$, $|L^i|=L$, we obtain the potential }
\be
V_{ \rm CP}(L)=  -A''(0)A''(L)\frac{1}{ 32 \pi^3} \frac{V''}{L^2} K_1(2 V'' L)\,.
\ee
This is a generalisation of the Casimir-Polder potential in the presence of an unscreened scalar, which matches results of  Refs.~\cite{Fichet:2017bng,Brax:2017xho} when taking
 $A(\phi)=\frac{\phi^2}{2M}$ and $V(\phi)=\frac{m^2}{2}\phi^2$.

The chameleon-like models are effective theories whose  predictions are valid below a cutoff scale, specific to each experimental situation.
 When self-interactions such as ${\cal L}\supset \phi^n/ \Lambda^{n-4}$ are present, the cutoff is expected to be  $\sim 4\pi \Lambda$ since higher order diagrams are expected to produce  fast-growing $1/(\Lambda L)^{n}$ contributions to the force which cannot be neglected when $L\sim 1/4\pi \Lambda$.
The cutoff resulting from the  interactions with matter is more subtle because of screening. Consider the contributions to the force from  the leading interaction
$M^{-2} \phi^2 J$ and  a next-to-leading interaction of order $M^{-4}\phi^4 J$ (shown in Fig.~\ref{fig:Diags}\textit{vi}), which contributes at two-loop as
\be
 {\rm 2-loop} \sim M^{-2}\int d^3x \partial_L J (\Delta(x,x))^2\, .
\ee
We obtain that the two loop contribution is negligible for~\footnote{
The two-loop contribution also renormalizes the leading order coupling. For instance in the case of plates studied in Sec.~\ref{sec:plate}, the two loop contribution is $(\Delta(L,L))^2-(\Delta(z_\infty,z_\infty))^2=
(\Delta^{\rm fin}(L,L))^2+2 \Delta^{\rm fin}(L,L) \Delta^{\rm fin}(z_\infty,z_\infty)$. The main divergence $(\Delta(z_\infty,z_\infty))^2$ cancels and the remaining divergence is absorbed in the renormalization of the leading coupling, leaving only the finite   term  $(\Delta^{\rm fin}(L,L))^2$.
}
\be
\Delta^{\rm fin}_J(x_{bd},x_{bd})\ll M^2 \,.
\label{eq:cutoff_matter}
 \ee
In the presence of screening, one has
\be
\Delta^{\rm fin}_J(x_{bd},x_{bd})\rightarrow 0 \,,
\ee
 see appendix \ref{app:A},  and therefore the  {\bf range of validity of the calculation of the quantum force is largely extended}. This is not surprising as the  Casimir pressure should not depend on the coupling to the plates, only on the mass and degrees of freedom of the field living between the plates. Also, this screening is a familiar effect in compact extra-dimension theories \cite{Brax:2004ym}:  a large brane mass term repels the field and amounts to a Dirichlet boundary condition \cite{Carena:2002me,Carena:2002dz}.
Yet, it is remarkable that the presence of screening  reduces the contributions from $n>1-$loop diagrams, hence improving on the $\hbar$ expansion.


Our conclusions about the validity of the chameleon-like EFT differ from those drawn in Ref.~\cite{Upadhye:2012vh} for the following reason. The reasoning of Ref.~\cite{Upadhye:2012vh} would  hold if the source occupied the whole space. However one should take into account that whenever an empty region exists, 
   the fluctuation gets confined  there when the effective mass induced by the source becomes large (as pictured in Fig.~\ref{fig:Diags}\textit{iv}).
 As a consequence the contributions to the $1-$loop  potential in the source region are suppressed by the vanishing wave function of the fluctuation, and the chameleon-like  EFT is not violated---even when the effective mass induced by  the source becomes infinite. 

\section{Quantum force between plates \label{sec:plate}}

{
We now study the case of a chameleon-like field in an environment whose constant density  changes piece-wise along the  direction $z$. We construct configurations with two facing plates, by first taking the plates to be of finite width to guarantee that the density vanishes at infinity and then taking the limit of infinite width.

The classical component of the chameleon force in this geometry has been extensively studied. It can be obtained using for instance  Eq.~\eqref{eq:Fclass} or  Eq.~\eqref{eq:claszbound}, which gives
$ F_{\rm cl} /S= \rho_3 \vert A(L) -A(\infty)\vert $, and can be further transformed to
\be
\frac{F_{\rm cl}}{S}= \vert V_J(\phi_{\rm vac}) -V_J(\phi_0)\vert
\ee
where $\phi_{\rm vac}$ is the values of the field  in the absence of the plates and $\phi_0$ is the value of the field midway between the plates, see \cite{Brax:2014zta}. We will recall the expressions of the classical pressure for inverse power-law chameleons \cite{Khoury:2003aq} and symmetrons \cite{Hinterbichler:2010es} in the next sections. The present section is about the quantum component force.

 We  model  the mass of the chameleon fluctuation  as a   piecewise constant along $z$. This piecewise mass model is important as  it is a sensible approximation whenever the 
profile of $\phi_{\rm cl}$ near the interfaces is irrelevant  compared the distance $L$.
This piecewise constant mass approximation is especially accurate for symmetron models~\cite{Brax:2017hna}. }


Let us then consider  three regions, for which the effective mass  $V_J''\equiv m^2(z)$ takes the form
\begin{align}   m^2(z)=    m_1^2 & \Theta(-z_\infty<z<0)+m_2^2 \Theta(0<z<L)
\nonumber \\
&+m_3^2 \Theta(L<z<z_\infty)   \,, \end{align}
where $|z_\infty|$ is near infinity, \textit{i.e.} larger than all other length scales of the problem.
This model can be readily used to calculate the chameleon pressure between plates of homogeneous mass density $\rho$,
in which case $m_2^2$ is seen as the intrinsic mass and the sources in regions 1, 3 are identified with
 $M^{-2} J_{1,3}= ( M)^{-2} \, \rho_{1,3}=m^2_{1,3}-m_2^2$.


Defining $\omega(z)=\sqrt{(p_0)^2-(p_1)^2-(p_2)^2+i \epsilon-m^2(z)}$, the equation of motion becomes
\be(\partial_z^2 + \omega^2(z)) \phi(z)=0\,.\ee
  The  solution in regions $i=1,2,3$ is simply $(\phi_i^+,\phi_i^-)=(e^{i\omega_i z},e^{-i\omega_i z})$. The solution everywhere   can be found by  continuity of the solution and its derivative at each of the  interfaces, defining momentum-dependent transfer matrices of the form
\be(\phi_2^+,\phi_2^-)^t=T_{21}(\phi_1^+,\phi_1^-)^t\,,\quad (\phi_3^+,\phi_3^-)^t=T_{32}T_{21}(\phi_1^+,\phi_1^-)^t \,.\ee
More details about the propagator are given in the Appendices \ref{app:A}, which also includes   details about the Feynman prescription and analytic continuation.

\begin{figure}
\includegraphics[width=8 cm,trim={0cm 0cm 0.cm 0cm},clip]{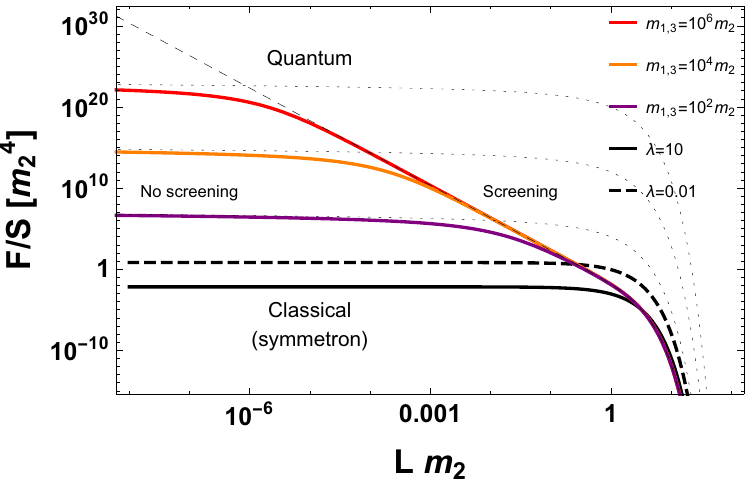}\\
\includegraphics[width=8 cm,trim={0cm 0cm 0.cm 0cm},clip]{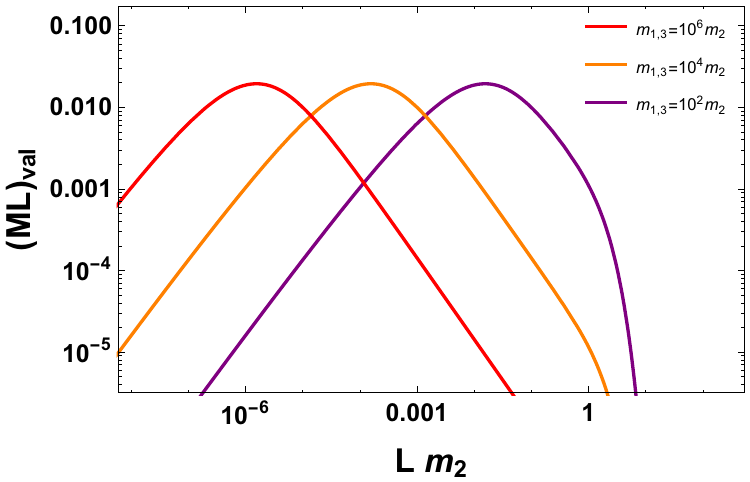}
\caption{Top: The quantum pressure between plates as a function of $L$ for fixed $m_2$ (or vice-versa).
Thin dashed line shows the Casimir pressure for a massive scalar.
Thin dotted lines show the integrated Casimir-Polder force.
The exact result interpolates between these two regimes.  The classical pressure
in the symmetron model where $m^2_2=2\mu^2$ is also shown. Bottom: Lower bound on  $M L$ needed for the perturbative expansion to be valid.
\label{fig:SymPlates}
}
\end{figure}

The quantum force induced by the fluctuation between regions 1 and 3 is obtained by varying $E[J]$ with respect to $L$, as described in Eq.~\eqref{eq:delE_full}. The variation of the source   gives  $  A'' \partial_L J =(m_2^2-m_3^2) \left(\delta(z-L) -\delta(z-z_\infty) \right)$.
{ This makes appear the quantity $\Delta_p(L,L)-\Delta_p(z_\infty,z_\infty) \equiv \Delta^{\rm fin}_p(L,L)$, where}
\be
\Delta_p(L,L)=
\frac{  (\omega_1+\omega_2)+  e^{2i L \omega_2} (\omega_2-\omega_1)
}{ (\omega_1+\omega_2)
(\omega_2+\omega_3)-
e^{2i L \omega_2} (\omega_2-\omega_1)(\omega_2-\omega_3)
}\,
\ee
and $\Delta_p(z_\infty,z_\infty)=1/(\omega_2+\omega_3)$.
 The final expression for the pressure between regions $1$ and $3$ is then
\begin{align}
&\frac{F_{\rm quant }}{S} =
 \int_0^\infty \frac{d \rho \rho^2}{2\pi^2} \\
& \frac{ \gamma_2 (\gamma_2-\gamma_1)(\gamma_2-\gamma_3)
}{ e^{2 L \gamma_2}
 (\gamma_1+\gamma_2)
(\gamma_2+\gamma_3)
-
 (\gamma_2-\gamma_1)(\gamma_2-\gamma_3) \nonumber
}\, \,
\label{eq:Density_cham}
\end{align}
where one has performed a Wick rotation and introduced $\omega_i=i\gamma_i=i \sqrt{\rho^2+m^2_i}$.

Let us consider some limiting cases. For $m_{1,3}\rightarrow \infty$, the expression gives the Casimir pressure from a massive scalar,
\be
\frac{F_{\rm quant }}{S} =  \int_0^\infty \frac{d \rho \rho^2}{2\pi^2}
 \frac{\gamma_2
}{
e^{2 L \gamma_2} -1
}\, \,
\ee
which is $\pi^2/(480L^4)$ if $m_2=0$.

On the other hand, weak coupling is defined by $(m^2_{1,3}-m_2^2)/ m^2_2\ll 1$ in which case a perturbative expansion is possible.  The leading order in the  expansion is quadratic and gives
\be
\frac{F_{\rm quant }}{S} = (m^2_{1}-m_2^2)(m^2_{3}-m_2^2)
\int_0^\infty \frac{d \rho \rho^2}{2\pi^2}
 \frac{e^{-2 L \gamma_2}
}{16
 (\gamma_2)^{3} }\,. \label{eq:CPint_direct}
\ee
This corresponds exactly to the Casimir-Polder force integrated over regions 1 and 3.

%
%

Although the limits taken above are conceptually simple, the transition between both as a function of $L$ is non trivial, as shown in Fig.~\ref{fig:SymPlates}. We see that the transition occurs over 3 orders of magnitude in $L$ and takes place near $L\sim 1/m_{1,3}$. Qualitatively, this is the typical distance for which the chameleon-like fluctuation has high enough momentum to start travelling in the $1,3$ regions.  This behaviour can be seen as a validity cutoff  on the  Casimir pressure, in the sense that at close enough distance  the pressure becomes constant instead of keeping growing. This behaviour can be see in Fig. \ref{fig:SymPlates}  at low values of $L$.  An $O(1)$ estimate of the pressure in this regime, taking $m_1 \sim m_3$, is given by
\be
\frac{F_{\rm quant}}{S} \sim - \frac{ (m^2_{1}-m_2^2)(m^2_{3}-m_2^2)}{32\pi^2} \log m_{1,3} L\,,
\label{appro}
\ee
and applies both for $m_{1,3}\gg m_2$ (strong coupling) and $m_{1,3} \sim m_2$ (weak coupling).

{The validity cutoff of the prediction of the quantum force in the presence of a higher-dimensional coupling to matter  is shown in
 Fig.~\ref{fig:SymPlates}. The minimum value allowed for $ML$, defined as
  $(M L)_{\rm val}\equiv L\sqrt{\Delta^{\rm fin}_J(x_{\rm bd},x_{\rm bd})}$
following Eq.~\eqref{eq:cutoff_matter}, reaches a maximal value of $\sim 0.02$, which is similar to $\sim1/4\pi$. This corresponds to the  \textit{lowest} possible  scale $M$ for a given $L$. Conversely, for a given $M$ this gives the domain of validity of the EFT as a function of $L$. 
 In the screening limit $m_{1,3}\gg 1/L$ the range of validity is largely widened. We have that $\Delta^{\rm fin}_J(x_{\rm bd},x_{\rm bd})\simeq \frac{\pi^2}{480 m_{1,3}^2 L^4}$ hence the EFT validity condition Eq.~\eqref{eq:cutoff_matter} takes the simple form
\be
L \gg \frac{1}{\sqrt{M m_{1,3}}}\,
\ee
which goes to much shorter length scales than $1/4\pi M$ .
}

\section{Quantum force in the E\"ot-Wash experiment}

\label{sec:eot}


The torsion pendulum E\"ot-Wash experiment involves two plates separated by a distance $L$ in which holes  have been drilled regularly on a circle. The two plates rotate with respect to each other. The scalar interaction induce a torque on the plates which depends on the potential energy of the configuration. The potential energy is obtained by calculating the amount of work required to approach one plate from infinity \cite{Brax:2008hh,Upadhye:2012qu}. Defining by $S(\theta)$ the surface area of the two plates which face each other at any given time, the torque is obtained as the derivative of the potential energy of the configuration with respect to the rotation angle $\theta$ and  is  given by
\begin{equation}
T \sim a_\theta \int_L^{\infty} dx \; \frac{ F}{S}(x)  \, ,
\end{equation}
where $a_\theta=\frac{dS}{d\theta}$ depends on the experiment.
For the 2006 E\"ot-Wash experiment~\cite{Kapner:2006si}, we consider the bound obtained
for a separation between the plates of $D=55\mu{\rm m}$,
\begin{equation}
\vert T \vert \le a_\theta \Lambda_T^3 ,
\end{equation}
where $\Lambda_T= 0.35 \Lambda $ \cite{Brax:2008hh} where $\Lambda\sim 2.4 $ meV is the dark energy scale.
Importantly, a thin electrostatic shielding sheet  is placed between the plates. It turns out that the presence of this sheet modifies both the classical and quantum components of the torque induced by the putative chameleon field, as we will see in the following.


The E\"ot-Wash experiment is sensitive to many modi\-fied gravity models, it is thus interesting to evaluate the  quantum force from a chameleon-like field in this setup. We use, as above, the piecewise constant mass approximation for the chameleon-like fluctuations.
Because of the electrostatic shielding  sheet present between the plates, the chameleon particle propagates in 5 different regions. The 5-layer propagator is obtained using the method described in Sec.~\ref{sec:plate}, and the subsequent change in force is shown in Fig.~\ref{fig:PlotEW} for a massless particle. We see that when the sheet is dense enough,
it screens the propagation and the quantum force is \textit{enhanced} by a factor 16, as the pressure is now between the plate and the sheet, twice closer than the opposite plate.
The full expression for the force in 5 regions is heavy and not very illuminating. But in the screening limit of the sheet the Casimir pressure is found to be $\pi^2/(30 (L-W_{\rm sheet} )^4)$, where $W_{\rm sheet}$ is the width of the sheet  ($L=55$\,$\mu$m, $W_{\rm sheet}=10$\,$\mu$m for \cite{Adelberger:2006dh}).

Interestingly,  without the sheet, the $L=55$\,$\mu$m E\"ot-Wash measurement is already close to be sensitive to the Casimir pressure  induced by a chameleon-like particle. Once the effect of the sheet is taken into account, the pressure is enhanced and E\"ot-Wash then becomes sensitive to the chameleon  Casimir pressure.


\section{Quantum bounds on chameleon-like models}

Here we consider the implications of the previous results  for two well-known models for which the piecewise constant mass approximation can be safely used.
In each model we first define the masses $m_i$ and give the classical force. The quantum force is obtained using the formalism developed in this paper.

\textit{Standard chameleon}.
The standard chameleon potential is
\be
V_J(\phi)= \frac{\Lambda^{4+n}}{\phi^n}+e^{\frac{\phi}{M}}J\,.
\ee
and has been studied in great details \cite{Burrage:2016bwy,Brax:2018zfb}.
The mass between the plates is given by
\be
m_2 L\approx  \sqrt{\frac{2(n+1)}{n}} B\left(\frac{1}{2}, \frac{1}{2} +\frac{1}{n}\right)
\ee
where $B$ is the Euler function. This is true as long as $L\gg m_{1,3}^{-1}$ where the masses in the plates are
$m_{1,3}^2= \Lambda^{n+4} / \phi_{1,3}^{n+2} $
and the fields in the plates are given by
$\phi_{1,3}^{n+1}= n\Lambda^{n+4}M J_{1,3}$.
The classical Casimir pressure is then
\be
\frac{F_{\rm cl}}{S}\approx  \Lambda^4 \left (\frac{\sqrt{2} n}{B(\frac{1}{2}, \frac{1}{2} +\frac{1}{n})}\Lambda L\right )^{-2n/(n+2)}
\ee
which is a power law as a function of $L$.

Focussing on $n=1$, the chameleon mass between the  plates is  $m_2\approx \pi /L$. We find that the most precise Casimir force experiments \cite{bordag2009advances,Decca:2007jq,Decca:2007yb} are sensitive to the small extra Casimir pressure induced by the chameleon.
It turns out that  exclusions from classical and quantum forces are complementary, and the quantum force excludes a large region inaccessible to other experiments, as shown in Fig.~\ref{fig:PlotEW}.

\textit{Symmetron}.
The symmetron model
 relies on the restoration of a $Z_2$ symmetry  in the presence of matter and is usually realised as
\be
V_J(\phi)=\frac{1}{2}\left(  \frac{1}{M^2} J -\mu^2 \right) \phi^2+\frac{\lambda}{4}\phi^4\,.\label{eq:Lsym}
\ee
When $J>\mu^2 M^2$ the expected value of the field is $\phi_{\rm cl}=0$ and the  mass of the fluctuation is $ \sqrt{M^{-2}J - \mu^2}$ whilst for smaller densities the field is  $\phi_{\rm cl}= \frac{\mu}{\sqrt \lambda}\equiv \phi_{\rm vac}$ and the  the mass is $\sqrt 2 \mu$. In the case of plates, one can show that the classical solution between plates vanishes when $ L\lesssim\pi/\mu$, in which case the classical pressure is given by
\be
\frac{F_{\rm cl}}{S}\approx \frac{\mu^4}{4\lambda} \,. \label{eq:Fsymcl}
\ee
When $ L\gtrsim\pi/\mu$ the classical solution between the plates increases until it reaches $\phi_{\rm vac}$ when $L\to \infty$.
The classical symmetron force between plates is suppressed and is approximated by
\be
\frac{F_{\rm cl}}{S}\approx \frac{\mu^4}{\lambda}e^{-2\sqrt 2\mu L}\, .
\ee
As made clear in Fig.~\ref{fig:SymPlates}, the classical force is  suppressed with respect to the quantum one by  $\sim (\mu L)^4/\lambda$ which is  small at distances $L<1/\mu$, for which the forces become active.

%
\begin{figure}
\includegraphics[width=7. cm,trim={0cm 0cm 0.cm 0cm},clip]{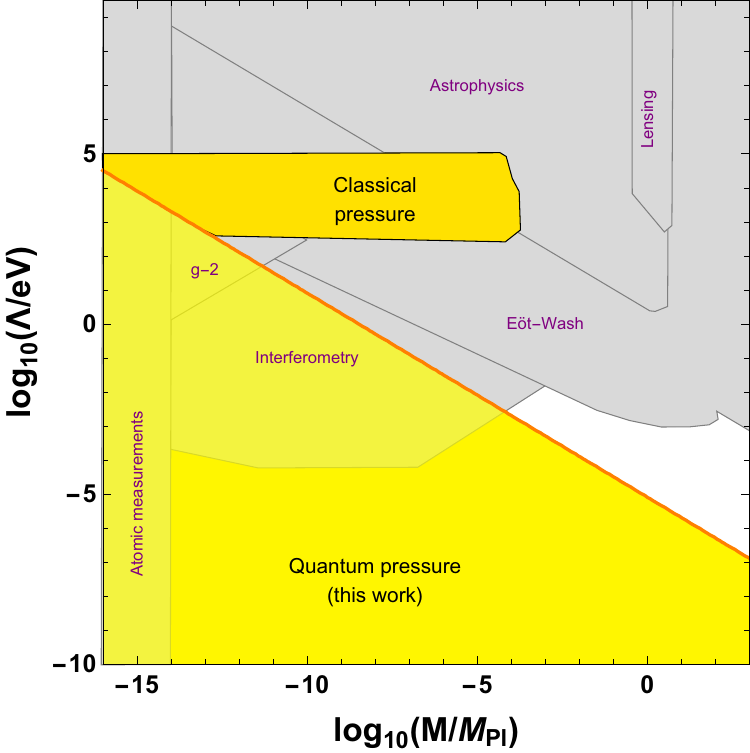}
\includegraphics[width=7. cm,trim={0cm 0cm 0.cm 0cm},clip]{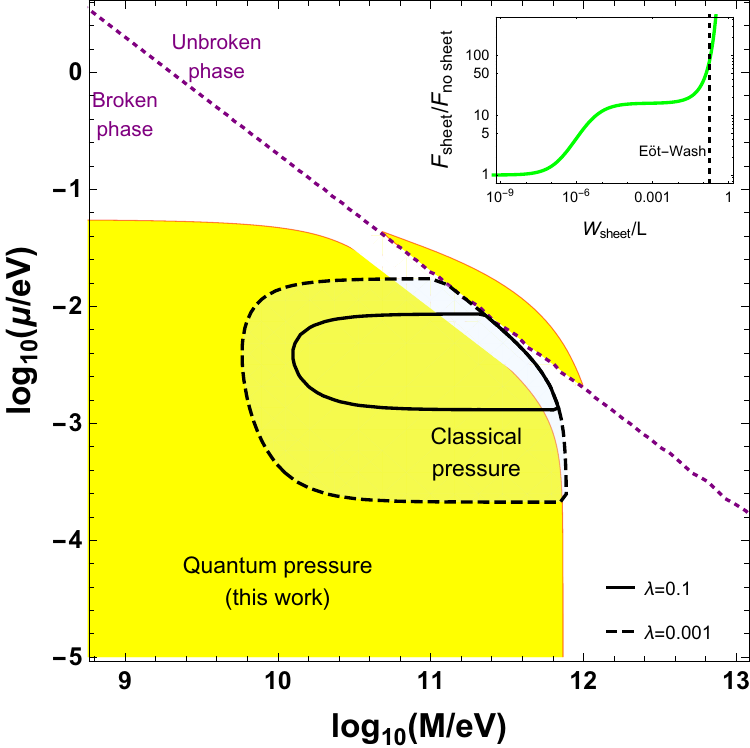}
\caption{
\label{fig:PlotEW}
Bounds on chameleon-like models. Top: Exclusion regions  on the standard chameleon with $n=1$, including bounds from Casimir experiments  on the quantum and classical chameleon pressures.
Bottom: Exclusion regions   on the symmetron  from E\"ot-Wash  on  quantum and classical  torques in presence of the intermediate sheet.
Insert: Enhancement of the quantum force from an intermediate sheet as a function of its width.
}
\end{figure}

A simple bound on the symmetron  comes from molecular spectroscopy, in which case the Casimir-Polder force between nuclei is unscreened and results from \cite{Fichet:2017bng,Brax:2017xho} can be applied.
For  masses below the meV range (see Fig.\,2 in  \cite{Brax:2017xho}), the main bound on the symmetron comes from the  E\"ot-Wash experiment.

Interestingly the E\"ot-Wash experiment is sensitive to the symmetron Casimir pressure \textit{because of} the  intermediate shield.  The $L=55\,\mu $m measurement  excludes a large part of the symmetron parameter space as shown in Fig.~\ref{fig:PlotEW}.
A sensitivity up to  $M\sim 1$\,TeV   and to $\mu\sim 58$\,meV  is obtained.  In comparison, the classical exclusion region \cite{Upadhye:2012rc,Brax:2014zta} is finite,   depends on $\lambda$ and vanishes for $\lambda\gtrsim 0.4$. The exclusion region near the transition requires a treatment of the VEV profile at the interface which is beyond the piecewise constant mass approximation used here.

Let us finally comment on the cosmological symmetron \cite{Hinterbichler:2010es,Hinterbichler:2011ca}.
In such case the parameters of the symmetron model  are typically chosen to  satisfy $\mu^2 M^2 \sim H_0^2 M_{\rm Pl}^2$, $\lambda \sim \frac{M_{\rm Pl}^2 \mu^2 }{M^4}$.
Solar system tests  require the coupling scale $M$  to be $M \lesssim 10^{-3} M_{\rm Pl}$.
It  turns out that the classical pressure Eq.~\eqref{eq:Fsymcl} is overwhelmed by the quantum pressure at the scale of laboratory experiments. The main constraint comes thus from the Eot-Wash  bound on the quantum pressure shown in Fig.~\ref{fig:PlotEW}. Since the lower bound on $M$ reaches only $ \sim 1$\,TeV, this leaves plenty of order of magnitudes in $M$ where  the cosmological symmetron can exist.
In terms of distance scales, the bound from the quantum pressure implies that the range of the symmetron force has to be \textit{larger} than $\sim 3\cdot 10^7 $~km.
 Simi\-lar conclusions apply to astrophysically relevant symmetrons, which tend to have larger masses $\mu$ and similar coupling scales $M$ \cite{OHare:2018ayv,Burrage:2018zuj}.

\section{Conclusion \label{se:conc}}

We have studied the forces induced by  chameleon-like particles in a fully-fledged quantum approach.
Our formalism elucidates the role of screening in the quantum picture and naturally interpolates between the limits of Casimir and Casimir-Polder pressures.
We have computed propagators with piecewise constant masses in an arbitrary number of 1D regions and analyzed in details the quantum chameleon pressure between plates. Our conclusions relative to the validity of the chameleon EFT  differ from \cite{Upadhye:2012vh} and are less restrictive.
In the E\"ot-Wash experiment we find that the sensitivity to the quantum pressure  from chameleon-like fields is enhanced by
 the presence of the intermediate sheet. For both symmetron and standard chameleon models, the bounds on the quantum pressure exclude large and previously unconstrained regions of the parameter spaces.


 \section*{Acknowledgements}
 SF thanks Orsay University for hospitality and funding.  This work is supported by the S\~ao Paulo Research Foundation (FAPESP) under grants \#2011/11973, \#2014/21477-2 and \#2018/11721-4.
 This work is supported in part by the EU Horizon 2020 research and innovation programme under the Marie-Sklodowska grant No. 690575. This article is based upon work related to the COST Action CA15117 (CANTATA) supported by COST (European Cooperation in Science and Technology).

\appendix

\section{Variation of the vacuum energy} \label{app:0}

Here we detail the one-loop calculation of the vacuum energy and its variation with the source term $J$. We start with the partition function
\be
Z[J]=\int {\cal D}\phi\, e^{i\int d^4x \left( \frac{1}{2}(\partial_\mu \phi)^2 -V(\phi)-A(\phi)J   \right)}\,.
\ee
Introducing $\phi=\phi_{\rm cl}+\eta$, where $\phi_{\rm cl}$ satisfies the classical equations of motion in the presence of the source $J$, gives
\be
Z[J]=\int {\cal D}\eta\, e^{i\int d^4x \left( \frac{1}{2}(\partial_\mu (\phi_{\rm cl}+\eta))^2 -V(\phi_{\rm cl}+\eta)-A(\phi_{\rm cl}+\eta)J   \right)}.
\ee
An infinitesimal change in the source $J\rightarrow J+\delta J$ as the one induced by the variation in $L$ also changes the classical field $\phi_{\rm cl}\rightarrow \phi_{\rm cl}+\delta\phi_{\rm cl}$. The partition function, once the source has been shifted, is explicitly given by
\begin{align}
 & Z[J+\delta J]=\int {\cal D}\eta\, \\
 & e^{-\int d^4x \left( \frac{1}{2}(\partial_\mu (\phi_{\rm cl}+\delta\phi_{\rm cl}+\eta))^2 -V(\phi_{\rm cl}+\delta\phi_{\rm cl}+\eta)-A(\phi_{\rm cl}+\delta\phi_{\rm cl}+\eta)(J+\delta J)   \right)} \,. \nonumber
\end{align}
However the shift in $\phi_{\rm cl}$ can be absorbed in the integration variable using $\tilde \eta \equiv \eta+\delta \phi_{\rm cl}$, leaving
\begin{align}
&Z[J+\delta J]=\\ & \int {\cal D}\tilde \eta\, e^{i\int d^4x \left( \frac{1}{2}(\partial_\mu (\phi_{\rm cl}+\tilde \eta))^2 -V(\phi_{\rm cl}+\tilde \eta)-A(\phi_{\rm cl}+\tilde \eta)(J+\delta J)   \right)}\,. \nonumber
\end{align}
Notice that the source variation $\delta J$ now only appears in the last term of the action. 
Let us  perform  the functional derivative of $E[J]$ built from this variation. This gives us the quantum average of $ \langle A \rangle_J$ as the variation only acts on the term in $ A(\phi_{\rm cl}+\eta) (J+\delta J)$
\be
\frac{E[J+\delta J]-E[J]}{\delta J}=
\frac{i}{Z[J]}\frac{Z[J+\delta J]-Z[J]}{\delta J}=  \langle A\rangle_J
\label{eq:EJDeltavar}
 \ee
in agreement with Eqs.~\eqref{eq:delE_simple}, \eqref{eq:Aav}.
Finally, on performs the $\eta$ integration of the $Z[J]$'s up to one-loop order, which is just the functional Gaussian integral
\begin{align}
Z[J]\approx & e^{i\int d^4x   \left( \frac{1}{2}(\partial_\mu \phi_{\rm cl})^2 -V(\phi_{\rm cl})-A(\phi_{\rm cl})J   \right)} \label{eq:ZJapprox1}
 \\ & \nonumber \cdot \left({\rm det}\left[\partial +V''+A'' J \right]\right)^{-1/2}\,, 
\end{align}
\begin{align}
Z[J+\delta J]\approx & e^{i\int d^4x   \left( \frac{1}{2}(\partial_\mu \phi_{\rm cl})^2 -V(\phi_{\rm cl})-A(\phi_{\rm cl})(J+\delta J)   \right)}
\label{eq:ZJapprox2}
 \\ & \nonumber \cdot \left({\rm det}\left[\partial +V''+A'' (J+\delta J) \right]\right)^{-1/2}\,. 
\end{align}
The first line in Eqs.~\eqref{eq:ZJapprox1},~\eqref{eq:ZJapprox2} contains the classical action whose variation gives the classical force.
The second line in these equations is the 1-loop functional determinant. Using these expressions in Eq.~\eqref{eq:EJDeltavar} gives the general formula Eq.~\eqref{eq:delE_full} of Sec.~\ref{sec:force} after some manipulations described in Eqs.~\eqref{eq:Trlog}-\eqref{eq:Deltageom}.

%

\section{Green's functions}
\label{app:A}

\subsection{Universal mass}

In position-momentum space, the Feynman propagator takes the form
\be
\Delta(z,z')=\frac{e^{i \omega |z-z'  |}}{2  \omega}\,
\ee
where one has introduced  \be\omega=\sqrt{(p_0)^2-(p_1)^2-(p_2)^2+i \epsilon-m^2} \,.\ee
This propagator can for instance be obtained by taking the Fourier transform of the usual 4-momentum space expression,
\begin{eqnarray}
&&\int \frac{dp_z}{2\pi} \frac{i}{p^2-m^2 +i \epsilon} e^{-i p_z (z-z')}\nonumber \\ && =
\int \frac{dp_z}{2\pi} \frac{-i}{p_z^2-\omega^2 -i \epsilon} e^{-i p_z (z-z')}\nonumber \\ &&= \frac{-i}{2\pi} \begin{cases}
(-2\pi i ) \frac{e^{i \omega (z-z')} }{-2\omega} \quad z>z' \\
(2\pi i ) \frac{e^{-i \omega (z-z')} }{2\omega} \quad z<z'
 \end{cases}\nonumber \\
\end{eqnarray}
We can see that, as a result of the $i\epsilon$ prescription, the propagator vanishes at infinity.
Note this is the boundary condition to impose if one calculates the position-momentum  propagator directly from the equation of motion.

\subsection{Piecewise constant mass }

Position-momentum space is convenient to treat the case of a $z$-dependent mass $m(z)$. Here we give the key steps to calculate the general case of $N$ regions ${\cal D}_i$,
\be
m^2(z)= \sum_{i=1}^N m^2_i \Theta (z\in {\cal D}_i)\,, \label{eq:massgen}
\ee
with  $\cup_{i=1}^N {\cal D}_i=\mathbf{R}$ and the interface between regions $j,j+1$ lies at the position $z_{j,j+1}$.
The Green's function satisfies
\be
\partial_z^2\Delta(z,z')-m^2(z)\Delta(z,z')=i \delta(z-z')\,.
\ee
The Feynman propagator is selected amongst the Green's function by imposing that it vanishes at infinity.
Defining $\omega(z)=\sqrt{(p_0)^2-(p_1)^2-(p_2)^2+i \epsilon-m^2(z)}$, the solutions in each region are given by $\phi^{\pm}(z)=e^{\pm i \omega_i z}$. Requiring continuity of the solution and of its derivative, the solution over the full space reads
 \be
\phi= \begin{pmatrix}
 A^+,  A^-
\end{pmatrix} \cdot \sum_{i=i}^N \left[ {\cal C}_i \begin{pmatrix}
\phi^+_i(x)
\\
\phi^-_i(x)
\end{pmatrix}\Theta(z\in {\cal D}_i )
\right] \label{eq:solgen}
 \ee
with
$
{\cal C}_i = \prod_{j=i}^{i}{\cal T}_{j, j+1}\,
$
where
\small
\begin{align}
&  {\cal T}_{j, j+1} = \frac{1}{2\gamma_{j+1}} \\ &
\begin{pmatrix}
(\gamma_j+\gamma_{j+1}) e^{i x_{j,{j+1}}(\gamma_j-\gamma_{j+1})} &
(\gamma_{j+1}-\gamma_j) e^{i x_{j,{j+1}}(\gamma_j+\gamma_{j+1})}
\\
(\gamma_{j+1}-\gamma_j) e^{-i x_{j,{j+1}}(\gamma_j+\gamma_{j+1})}
 &
(\gamma_j+\gamma_{j+1}) e^{i x_{j,{j+1}}(\gamma_{j+1}-\gamma_{j})}
 \end{pmatrix} \nonumber
\end{align}
\normalsize
is the transfer matrix given by the continuity conditions. The rest of the calculation of the Green's function is given by standard ODE solving techniques,
 see for instance the appendix of \cite{mypaper} for more details.

\subsection{The two-regions case }

In case of two regions, the Feynman propagator is found to be
\be
\Delta(z, z')= \label{eq:propa2regions}
\begin{cases}
 \frac{e^{i\omega_2 (z_>-z_<)}}{2\omega_2}E_2(z_<)   \quad\quad z_{12}<z_<
 \\
 \frac{e^{i(\omega_2 (z_>-z_{12})-\omega_1 (z_<-z_{12})}}{\omega_1+\omega_2}   \quad z_<<z_{12}< z_>
  \\
  \frac{e^{i\omega_1 (z_>-z_<)}}{2\omega_1}E_1(z_>)  \quad\quad  z_><z_{12}
\end{cases}\,
\ee
where
\begin{eqnarray}
&& E_1(z)=1+ e^{i2(z_{12}-z)\omega_1}\frac{\omega_1-\omega_2}{\omega_1+\omega_2}\nonumber \\ && E_2(z)=1+ e^{i2(z-z_{12})\omega_2}\frac{\omega_2-\omega_1}{\omega_1+\omega_2}\,
\nonumber \\
\end{eqnarray}
and  where $x_<= \min (x,x')$ and $x_>=\max (x,x')$.
The $E_1$, $E_2$ functions essentially describe how the presence of the boundary affects the propagator with both endpoints in the same region. When the boundary $x_{12}$ is rejected to infinity, one recovers the usual expression for  fully homogeneous space.

\section{Vanishing at the boundary}

Let us consider two regions of arbitrary shape ${\cal D}_1$, ${\cal D}_2$ where the mass takes values $m_1$, $m_2$. When $m_2\rightarrow \infty$ while other scales remain fixed, the homogeneous  equation of motion in ${\cal D}_2$ corresponds to  $m_2^2 \Phi=0$ which is satisfied only if $\Phi=0$ at any point in ${\cal D}_2$. Moreover, since one requires continuity of the solution in the whole space, the value of $\Phi$ at the interface is also set to zero when  $m_2\rightarrow \infty$. As a result, the problem is equivalent to having a field living in ${\cal D}_1$ and  a Dirichlet boundary condition at the boundary of ${\cal D}_2$.  This property can be directly seen in the planar case in, for instance, Eq.~\eqref{eq:propa2regions}. For $\omega_2\rightarrow \infty$, it is clear that the propagator tends to zero inside ${\cal D}_2$ and at its boundary.

\section{Analytic structure}

The quantum force at one loop has been calculated in Sec.~\ref{sec:plate} using a Wick rotation in $p_0$.

Let us first verify that the integrand $\Delta_p(L,L)$ (and $\Delta_p(z_\infty,z_\infty)$) are analytic in the first and third quadrant of the $p_0$ complex plane.  The function $\Delta_p(L,L)$ depends on the $w_i's$ which have branch cuts on intervals along the real axis. The $i\epsilon$ prescription shifts the branch cuts just below the real axis for $p_0>0$ and just above the real axis for $p_0<0$. Thus the integrals in the $p_0$ plane along the real axis avoid the branch cuts, and no branch cut is crossed during the Wick rotation.
  Let us analyse the poles of the integrand. They would appear for
\be
\frac{\omega_1-\omega_2}{\omega_1 +\omega_2} \frac{\omega_2-\omega_3}{\omega_2 +\omega_3}= e^{-2i\omega_2 x_0}.
\ee
In the first quadrant of the $p_0$ complex plane and writing $\omega_i= \vert \omega_i\vert e^{i\theta_i}$ we have $\theta_{1,3} > \theta_2$ as $m_{1,3}>m_2$. These angles are all in the first quadrant too. This implies that
\be
\left| \frac{\omega_1-\omega_2}{\omega_1 +\omega_2} \frac{\omega_2-\omega_3}{\omega_2 +\omega_3}\right| <1
\ee
whilst $\vert e^{-2i\omega_2 x_0}\vert = 1$. Hence the integrand has no poles in the first quadrant of the complex $p_0$ plane and one can perform a Wick's rotation to the imaginary axis. A similar analysis applies to the third quadrant.

Finally, the $\Delta_p(L,L)-\Delta_p(z_\infty,z_\infty)$ integrand tends exponentially to zero at infinity on the arcs in the first and third quadrant, including on the real axis because of the $i\epsilon$ shift, hereby ensuing that the Wick rotation is valid just like in the familiar case of a universal mass.

\section{The Casimir-Polder force}
\label{app:B}
In the main text, the Casimir-Polder force between plates (\ref{eq:CPint_direct})   has been obtained as the unscreened limit of the general result  Eq.~(11), which is given by the path integral approach introduced in this work.
Here we present an alternative calculation of the Casimir-Polder force between plates, done by first calculating the Casimir-Polder force between point-like sources using the Feynman diagram approach and then integrating over the plates. The result matches the unscreened limit   (\ref{eq:CPint_direct})  obtained in the main text.

 Rewrite the source term as
\begin{eqnarray}
&&{\cal L} \supset A'' J(x)= \frac{1}{2} m_2^2 \eta^2 +
\frac{1}{2} \eta^2 ( \Theta(x<0) (m_1^2 -m_2^2)\nonumber \\
&& +   \Theta(r<x) (m_3^2-m_2^2)). \nonumber \\
\end{eqnarray}
We consider the presence of the plates as small perturbations, related to the coupling to individual nucleons via
\be
(m_i^2 -m_2^2)=\frac{\rho_i}{\Lambda^2}=\frac{m_N n_i}{\Lambda^2}=\frac{m_N N_i}{V_i\Lambda^2}
\ee
where $\rho_i$ is the mass density, $n_i$ is the number density, $N_i$ is the total number of particles homogeneously distributed in the volume $V_i$.

We first compute the potential between two point sources (the single static nucleons), replacing $\rho$ by $m_N \delta^{(3)}(x)$. The corresponding source term is
\be
{\cal L} \supset A J(x)=\frac{1}{2} \eta^2
\left(
 \frac{m_N}{\Lambda^2} \delta^{(3)}(x^i-x^i_a) +   \frac{m_N}{\Lambda^2} \delta^{(3)}(x^i-x^i_b) \right)\,.
\ee
The bubble diagram  is
\be
i{\cal M} = - \frac{m_N^2}{\Lambda^4} \,4 m_N^2\, \frac{1}{2} \int \frac{dk^3}{(2\pi)^3} \frac{e^{i\omega_2|z_1-z_2|)}}{2\omega_2}
\frac{e^{i\omega'_2|z_1-z_2|)}}{2\omega'_2}
\ee
where $\omega_2=\sqrt{k^2-m_2^2}$, $\omega'_2=\sqrt{(k+p)^2-m_2^2}$. We have used the explicit expression for the Feynman propagator.
In this formalism $k,p$ are 3-momenta, $k=(k^0,k^1,k^2)$ for example.
The scattering potential is given by
\begin{eqnarray}
&&\tilde V(p,z_1-z_2) = - \frac{\cal{M}}{4 m_N^2}\nonumber \\
&&= -i \frac{m_N^2}{\Lambda^4} \frac{1}{2}  \int \frac{dk^3}{(2\pi)^3} \frac{e^{i\gamma_2|z_1-z_2|)}}{2\omega_2}
\frac{e^{i\omega_2'|z_1-z_2|)}}{2\omega_2'}\,.\nonumber \\
\end{eqnarray}
The sources are static hence $p_0$ can readily be set to zero.
The spatial potential is given by the Fourier transform of this,
\be
V\left(\sqrt{(z_1-z_2)^2+x_\parallel^2}\right)= \int \frac{d^2 p_{\parallel} }{(2\pi)^2} \tilde V(p_{\parallel},z_1,z_2)e^{i p_{\parallel}\cdot x_{\parallel} }
\ee
where $x_\parallel=(x^1,x^2 )$.
We are also going to average the potential over plates with separation $L$,
\begin{eqnarray}
&&V_{\rm plates}= N_1 V_1^{-1}N_3V_3^{-1} \times \nonumber \\ && \int d^2x_{\parallel}\int_{-\infty}^0 dz_1  \int d^2x'_{\parallel}\int^{\infty}_{L} dz_2
 \int \frac{d^2 p_{\parallel} }{(2\pi)^2} \tilde V(p_{\parallel},z_1,z_2)e^{i p_{\parallel} \cdot x_{\parallel} }\nonumber \\
\end{eqnarray}
where $V_1$, $V_3$ are the volumes of regions 1 and 3.
One has $\int d^2x'_{\parallel}=S$.
We can see that the integrals simplify since
\be
\int d^2x_{\parallel} \frac{d^2 p_{\parallel} }{(2\pi)^2}  e^{i p_{\parallel}\cdot x_{\parallel} } F(p_{\parallel}) =F(0)\,.
\ee
Thus the potential is simply
\begin{align}
V_{\rm plates}&= S n_1n_3 \int_{-\infty}^0 dz_1  \int^{\infty}_{L} dz_2    \tilde V(0,z_1,z_2) \\
&= - i S n_1n_3\frac{m_N^2}{\Lambda^4} \frac{1}{2} \int \frac{dk^3}{(2\pi)^3}
\int_{-\infty}^0 dz_1  \int^{\infty}_{L} dz_2
 \frac{e^{i2\omega_2(z_2-z_1)}}{4\omega^2_2} \\
 &=  S n_1n_3 \frac{m_N^2}{\Lambda^4}  \, \int \frac{dk_E^3}{(2\pi)^3}
\int_{-\infty}^0 dz_1  \int^{\infty}_{L} dz_2
 \frac{e^{-2\gamma_2(z_2-z_1)}}{-8(\gamma_2)^2} \\
 &= - S n_1n_3 \frac{m_N^2}{\Lambda^4}
 \, \int \frac{dk_E^3}{(2\pi)^3}
 \frac{e^{-2\gamma_2 L }}{32(\gamma_2)^4} \\
  &= - S (m_1^2-m_2^2 ) (m_3^2-m_2^2 )  \frac{m_N^2}{\Lambda^4}
 \, \int \frac{dk_E^3}{(2\pi)^3}
 \frac{e^{-2\gamma_2 L }}{32(\gamma_2)^4}\,.
\end{align}
In the last line one has used $\frac{n_i\,m_N}{\Lambda^2}=m_i^2-m_2^2$. 
Finally the pressure is obtained by taking the  derivative
\be
P=S^{-1}\partial_L V_{\rm plates}=(m_1^2-m_2^2 ) (m_3^2-m_2^2 )
 \, \int \frac{dk_E^3}{(2\pi)^3}
 \frac{e^{-2\gamma_2 L }}{16(\gamma_2)^3}\,.
\ee
This reproduces (\ref{eq:CPint_direct} ) in the main text.

\section*{References}

\bibliography{biblio}

\end{document}